\documentclass[preprintnumbers,amsmath,amssymb]{revtex4}

\usepackage{graphics}
\usepackage{epsfig}
\usepackage{dcolumn}
\usepackage{bm}

\begin{document}

\vspace*{1.0cm}

\begin{center}
{\Large {\bf PQCD Calculation for $\Lambda_{b}\rightarrow \Lambda
\gamma$ in the Standard Model}}
\end{center}
\vspace*{0.3cm}
\begin{center}
 { Xiao-Gang He$^{1,2}$, Tong Li$^1$, Xue-Qian Li$^1$, and Yu-Ming Wang$^1$}\\
\vspace*{0.3cm}
$^1$Department of Physics, Nankai University, Tianjin\\

\vspace*{0.1in}
 $^2$Center for Theoretical Sciences, Department of Physics,
 National Taiwan University,
 Taipei\\
\end{center}

 \vspace*{0.5cm}

\begin{center}
\begin{minipage}{12cm}

\noindent Abstract:\\

We calculate the branching ratio of $\Lambda_b \to \Lambda \gamma$
in the standard model using the PQCD method. The predicted
branching ratio $B(\Lambda_b \to \Lambda \gamma)$ is about
$(4.3\sim 8.6)\times 10^{-8}$, with reasonable parameter ranges in
the heavy baryon distribution amplitude. This branching ratio is
much smaller than those obtained in other hadronic model
calculations. Future experimental data can provide important
information on applicability of the PQCD method to heavy baryon
radiative decay.
\end{minipage}

\end{center}

\vspace{2 cm} PACS numbers: 13.30.Ce, 12.38.Bx, 14.20.Mr

\newpage

\section{Introduction}
Rare radiative processes involving $b\to s \gamma$ at quark level
are important for understanding the flavor changing structure in
the standard model (SM). Exclusive radiative $B$ decays also
provide important information about the hadronic matrix elements
where a heavy b-quark is involved. These processes being rare can
also provide clues to models beyond the SM. There have been
considerable studies on inclusive $b\to s \gamma$\cite{CLEO,b to s
gamma exp}, and exclusive mesonic $B\to K^* \gamma$\cite{exclusive
gamma exp} both experimentally and theoretically within and beyond
the SM \cite{b to s gamma th,exclusive gamma th,newbgamma}.
Theoretical predictions for inclusive decays agree with data very
well in the SM. Calculations for exclusive processes are in
general consistent with data although there are unavoidable
uncertainties due to our lack of good understanding of QCD at low
energies. Nevertheless methods have been developed to calculate
hadronic matrix elements in recent years
\cite{hadronmesonBeneke,hadronmesonLi}. With more data becoming
available, new b-decay processes can be studied. These processes
can be new tests for different methods in calculating hadronic
matrix elements and new physics beyond the SM. In this work we
study $\Lambda_b\to \Lambda \gamma$. In this decay more
experimental information about the heavy b quark inside the hadron
which is not available in inclusive and mesonic b-hadron decays,
such as spin polarization during hadronization, and the handedness
of the couplings at the quark level, can be
extracted\cite{llll,Mannel,Huang,Mohanta,Cheng}. Therefore the
baryonic b-hadron radiative decay can provide a new test for
theoretical methods for b-quark hadronization.

There are some studies in the literature on $\Lambda_b \to \Lambda
\gamma$\cite{llll,Mannel,Huang,Mohanta,Cheng} decay ranging from
phenomenological models to QCD sum rule approaches. Our study will
be based on the PQCD method\cite{Lip,Lic,LiJ}. This method has
been shown to give consistent results for two body mesonic B
decays\cite{hadronmesonLi}. We expect a PQCD calculation for
$\Lambda_b \to \Lambda \gamma$ will also give a reasonable
estimate since the energy-exchange carried by gluons in the matrix
element calculations is large. Result obtained in this way can
serve as a good reference for discussing the relevant hadronic
matrix elements.

For SM, the effective Hamiltonian responsible for $b\to s\gamma$
comes from the electromagnetic penguin diagram and is given
by\cite{H}:
\begin{eqnarray}
H_{eff} &=& i \frac{G_{F}}{2\sqrt{2}}V_{tb}V_{ts}^{*} {e\over
4\pi^2} C^{eff}_7(\mu)m_b
\bar{s}\sigma_{\mu\nu}(1+\gamma_5)bF^{\mu\nu}, \label{gamma}
\end{eqnarray}
where $C_7^{eff}(\mu = m_b) = -0.31$. In our numerical
calculations, the running of $C_7^{eff}$ will also be taken into
account.

It has been shown that there may be resonant (long distance)
$J/\psi (\psi')$ contributions\cite{long0}. If these contributions
are included, one should add a term $
(3C_1(\mu)+C_2(\mu))(3/\alpha_{em}^{2})\sum_{j=\psi,\psi'}
\omega_j(0)k_j\pi\Gamma(j\rightarrow
l^{+}l^{-})M_{j}/(q^2-M_j^2+iM_{j}\Gamma_j^{tot})$ to the Wilson
coefficient $C^{eff}_7$. Since for $b\to s \gamma$ process,
$q^2=0$, there are double suppressions for the long distance
resonant contributions with one of them coming from the
Breit-Wigner factor $\sim \Gamma_i/M_i$ and another coming from
the extrapolation of $\omega_j(M^2_i)=1$ to $\omega_j(0)$ with
$\omega_j(0) <0.13$ (and could be smaller)\cite{long0}, we will
neglect the resonant contribution for radiative decays in our
later discussions.

At the hadron level, the decay amplitude for $\Lambda_b\to \Lambda
\gamma$ is obtained by inserting the effective Hamiltonian between
the initial and final hadron states,
\begin{eqnarray}
M(\Lambda_b \to \Lambda \gamma) = \langle \Lambda \gamma\vert
H_{eff}\vert \Lambda_b\rangle.
\end{eqnarray}

There are two form factors for $\Lambda_b \to \Lambda \gamma$ from
the above which we write as
\begin{eqnarray}
M_{\mu} &&\equiv
\langle\Lambda(p')|C_7^{eff}(\mu,0)\bar{s}\sigma_{\mu \nu}
q^{\nu}(1+\gamma_5) b|\Lambda_{b}(p)\rangle  =
\overline{\Lambda(p')}(F_L\sigma_{\mu \nu}
q^{\nu}(1-\gamma_5)+F_R\sigma_{\mu \nu}
q^{\nu}(1+\gamma_5))\Lambda_{b}(p).\label{formfactor}
\end{eqnarray}

We obtain
\begin{eqnarray}
\Gamma(\Lambda_b \to \Lambda \gamma) =
{G_F^2|V_{tb}|^2|V_{ts}|^2\alpha_{em}|C_7^{eff}|^2 m_b^2\over 32
m_{\Lambda_b}^3\pi^4}(m_{\Lambda_b}^2-m_{\Lambda}^2)^3(|F_L|^2 +
|F_R|^2).
\end{eqnarray}

Emission of a photon from the tree operators $O_{1,2}$ can also
contribute to $\Lambda_b \to \Lambda \gamma$. Although the Wilson
coefficients of these operators are larger than those of the
penguin operators, there is a large suppression coming from the
CKM factor $|V_{ub}V_{us}^*/V_{tb}V_{ts}^*|$. The overall
contributions from bremsstrahlung of a photon off the operator
$O_{1,2}$ is therefore suppressed. We will neglect their
contribution in rest of discussions.

\section{PQCD calculation of the hadronic matrix elements}

We now describe our calculations for the hadronic matrix elements
defined above using the PQCD method developed in
Ref.\cite{Lip,Lic,LiJ}. We define, in the rest frame of
$\Lambda_{b}$,  $p$, $p'$ to be the $\Lambda_{b}$, $\Lambda$
momenta, $k_i(i=1,2 ,3)$ to be the valence quark momenta inside
$\Lambda_{b}$, and $k_i'$ to be the valence quark momenta inside
$\Lambda$. We parameterize the light cone momenta with all light
quark and baryon masses neglected as
\begin{eqnarray}
&&p = (p^{+},p^{-},\mathbf{0}_{T}) = \frac
{M_{\Lambda_{b}}}{\sqrt{2}}(1,1,\mathbf{0}_{T}),
\;\;p' = (p'^{+},0,\mathbf{0}_{T})\nonumber \\
&&k_{1} = (p^{+},x_1p^{-},\textbf{k}_{1T}),\;\;k_{2} =
(0,x_2p^{-},\textbf{k}_{2T}),
\;\;k_{3} = (0,x_3p^{-},\textbf{k}_{3T})\nonumber \\
&&k'_{1} = (x_{1}'p'^{+},0,\textbf{k}_{1T}'),\;\;k'_{2} =
(x_{2}'p'^{+},0,\textbf{k}_{2T}'),\;\;k'_{3} =
(x_{3}'p'^{+},0,\textbf{k}_{3T}')
\end{eqnarray}
where $x_i$ and $x'_i$ are the fractions of the longitudinal
momenta of the valence quarks with $x_1+x_2+x_3 = 1$ and
$x'_1+x'_2+x'_3 = 1$. $\mathbf{k}_{iT}$ and $\mathbf{k}_{iT}'$ are
the transverse momenta of the valence quarks inside $\Lambda_b$
and $\Lambda$, respectively.

As a self-consistent check, one should make sure that the expected
relation $p^2 - k_1^2 \sim 0(\Lambda_{QCD m_b})$ holds, since the
light quarks in the heavy baryon should have momenta of order
$\Lambda_{QCD}$. Naively, the above gives a value of order
$(1-x_2)m^2_{\Lambda_b}$ which does not have the explicit form as
expected. To understand this, one needs to combine the form of the
heavy baryon wavefunction which determines how quark momenta are
distributed inside the baryon. We have checked this using the
wavefunction given later, and obtained the ratio of average values
$<x_{2,3}>/<x_1> \sim m/m_{\Lambda_b}$, where $m$ is of order
$\Lambda_{QCD}$. With the constraint $x_1+x_2+x_3 = 1$, the
desired order for $p^2-k^2_1$ is then obtained.

One can write $p'^+=\rho p^{+}$ with $\rho=\frac{2 p \cdot
p'}{M_{\Lambda_{b}}^{2}} ={p^2+p'^2-q^2\over M_{\Lambda_b}^2}$.
The ranges for $q^2$ and $\rho$ are given by $0 \leq q^2 \leq
(M_{\Lambda_b}-m_\Lambda)^2$ and $ {2m_\Lambda/M_{\Lambda_b}}\leq
\rho \leq ({M_{\Lambda_b}^2+m_{\Lambda}^2)/ M_{\Lambda_b}^2}$ if
off-shell photon is allowed. In our case of $\Lambda_b \to \Lambda
\gamma$, $q^2=0$. Here we have kept $\Lambda$ mass in the
expressions for the purpose in tracing the ranges of the kinematic
variables. In the approximation we are using, it should be set to
zero as mentioned above.

In the PQCD picture, hadrons are formed from quarks with
appropriate wave functions describing the momenta distribution of
quarks inside the hadron. The $\Lambda_b$ wave function is usually
defined through the quantity \cite{lambdabwave,new1}.
\begin{eqnarray}
&&(Y_{\Lambda_b})_{\alpha \beta \gamma}(k_{i},\nu) =
\frac{1}{2\sqrt{2}N_{c}}
 \int \prod_{l=2}^{3}\frac{d w_{l}^{+} d \bf{w}_l }{(2\pi)^{3}} e^{ik_{l}w_{l}}
 \varepsilon^{abc}\langle0|T[b_{\alpha}^{a}(0)u_{\beta}^{b}(w_{2})d_{\gamma}^{c}(w_{3})]|
 \Lambda_b(p)\rangle\nonumber \\
&&=\frac{f_{\Lambda_b}}{8\sqrt{2}N_{c}}[(p\!\!\! \slash
+M_{\Lambda_b})\gamma_{5}C]_{\beta \gamma}
[\Lambda_b(p)]_{\alpha}\Psi(k_{i},\nu), \label{YYb}
\end{eqnarray}
where $f_{\Lambda_b}$ is a normalization constant, $\Lambda_b(p)$
is the $\Lambda_b$ spinor, and $\Psi(k_i, \mu)$ is the wave
function. Here we have used the heavy quark symmetry which should
be applicable in the present case, following
Refs.\cite{lambdabwave,new1}, to reduce the form factors to the
above simplified form. In general there are more components in the
wavefunction if all quarks are light. For the light baryon
$\Lambda$ the leading-twist wave function of $\Lambda$ is defined
by\cite{lambdawave}:
\begin{eqnarray}
&&(Y_{\Lambda})_{\alpha \beta \gamma}(k_{i}',\nu) =
\frac{1}{2\sqrt{2}N_{c}} \int \prod_{l=1}^{2}\frac{d w_{l}^{-} d
\mathbf{w_{l}} }{(2\pi)^{3}} e^{ik_{l}'w_{l}}
\varepsilon^{abc}\langle0|T[s_{\alpha}^{a}(w_{1})
u_{\beta}^{b}(w_{2})d_{\gamma}^{c}(0)]|
\Lambda(p')\rangle \nonumber \\
&&=\frac{f_{\Lambda}}{8\sqrt{2}N_{c}}\{(p'\!\!\! \slash C)_{
\beta\gamma} [\gamma_{5}\Lambda(p')]_{\alpha}\Phi^{V}(k_{i}',\nu)
+(p'\!\!\! \slash \gamma_{5} C)_{\beta\gamma}[\Lambda(p')]_{\alpha}\Phi^{A}(k_{i}',\nu)\}\nonumber \\
&&-\frac{f_{\Lambda}^{T}}{8\sqrt{2}N_{c}}(\sigma_{\mu
\nu}p'^{\nu}C)_{\beta\gamma}
[\gamma^{\mu}\gamma_{5}\Lambda(p')]_{\alpha}\Phi^{T}(k_{i}',\nu),
\label{YY}
\end{eqnarray}
where $f_{\Lambda}$ and $f_{\Lambda}^T$ are normalization
constants, and $\Lambda(p')$ is the $\Lambda$ spinor.

Including the Sudakov factor with infrared cut-offs
$\omega(\omega')$, and running the wavefunction from $\nu$ down to
$\omega (\omega')$, then we obtain\cite{Lip}:
\begin{eqnarray}
&&\Psi(x_{i},b_{i},p,\nu)=\mathrm{exp}[-\sum_{l=2}^{3}s(\omega,x_lp^-)-3\int_{\omega}^{\nu}
\frac{d
\overline{\mu}}{\overline{\mu}}\gamma_q(\alpha_{s}(\overline{\mu}))
]\Psi(x_{i})\nonumber \\
&&\Phi^{j}(x_{i}',b_{i}',p',\nu)=\mathrm{exp}[-\sum_{l=1}^{3}s(\omega',x_{l}'p^{+})-3\int_{\omega'}^{\nu}
\frac{d\overline{\mu}}{\overline{\mu}}\gamma_q(\alpha_{s}(\overline{\mu}))]
\Phi^{j}(x_{i}'),\nonumber \\
\label{suda}
\end{eqnarray}
where $j = V,\; A,\; T$, $\omega = min(1/\tilde b_1,1/\tilde
b_2,1/\tilde b_3)$, and $\omega' = min(1/\tilde b'_1,1/\tilde
b'_2,1/\tilde b'_3)$. $\tilde b^{(')}_1
=|\textbf{b}^{(')}_2-\textbf{b}^{(')}_3|$, $\tilde b^{(')}_2
=|\textbf{b}^{(')}_1-\textbf{b}^{(')}_3|$, and $\tilde b^{(')}_3
=|\textbf{b}^{(')}_1-\textbf{b}^{(')}_2|$. Here $\textbf{b}$ and
$\textbf{b}'$ are the conjugate variables to $\textbf{k}_T$ and
$\textbf{k}'_{T}$ defined in Appendix B.

The explicit expressions for the Sudakov factors are given in
Ref.\cite{Lip} with
\begin{eqnarray}
&&s(\omega,Q)=\int^Q_\omega{dp\over p}[ln({Q\over
p})A[\alpha_s(p)]+B[\alpha_s(p)]]\nonumber \\
&&A=C_F{\alpha_s\over \pi}+[{67\over 9}-{\pi^2\over 3}-{10\over
27}n_f+{8\over 3}\beta_0ln({e^{\gamma_E}\over 2})]({\alpha_s\over
\pi})^2\nonumber \\
&&B={2\over 3}{\alpha_s\over \pi}ln({e^{2\gamma_E-1}\over
2})\nonumber \\
&&\gamma_q (\alpha_s(\mu)) = -\alpha_s(\mu)/\pi,\nonumber\\
&&\beta_0={33-2n_f\over 12},
\end{eqnarray}
where $\gamma_E$ is the Euler constant. $n_f$ is the flavor
number, and $\gamma_q$ is the anomalous dimension. For $\Lambda_b$
baryon decays, the typical energy scale is above the charm mass.
We will take $n_f$ equal to 4 in our calculations.

The hadronic matrix elements can be written as:
\begin{eqnarray}
&&M_{l,\mu}= \int[Dx]\int[Db](\overline{Y}_{\Lambda})_{\alpha'
\beta'\gamma'}(x_i',b_i',p',\nu)\nonumber \\
&&H_{l,\mu}^{\alpha'\beta'\gamma'\alpha\beta\gamma}(x_i,x_i',b_i,b_i',M_{\Lambda_b},\nu)
(Y_{\Lambda_b})_{\alpha\beta\gamma}(x_i,b_i,p,\nu),
\label{integral}
\end{eqnarray}
where the measures of the momentum fractions \cite{Lip} are give
by
\begin{eqnarray}
&&[Dx] = [dx][dx'],\;\;[dx] = dx_1 dx_2 dx_3
\delta(1-\sum_{l=1}^{3}x_l),\;\;[dx'] =
dx_{1}'dx_{2}'dx_3'\delta(1-\sum_{l=1}^{3}x_l').
\end{eqnarray}
The measures of the transverse extents $[Db]$ are defined in
Appendix A.

The hard scattering amplitude
$H_{l,\mu}^{\alpha'\beta'\gamma'\alpha \beta
\gamma}(x,x',b,b',M_{\Lambda_b},\nu)$ is obtained by first
evaluating the amplitude
$H_{l,\mu}^{i,\alpha'\beta'\gamma'\alpha\beta\gamma}(x_i,x_i',\textbf{k}_{T},\textbf{k}_{T}'
,M_{\Lambda_b})$ for the 'i'th diagram in Fig. \ref{diagrams} for
a corresponding Wilson coefficient $C_l^{eff}$ which is displayed
in Appendix B. One then carries out a Fourier transformation on
$\textbf{k}_T$ and $\textbf{k}'_T$ to $\vec b$ and $\vec b'$ space
to obtain $\tilde H_{l,\mu}^{i,\alpha'\beta'\gamma'\alpha \beta
\gamma}(x,x',b,b',M_{\Lambda_b})$. The procedure of carrying out
this transformation is described at the end of Appendix B.

\begin{figure}[!htb]
\begin{center}
\begin{tabular}{cc}
\includegraphics[width=16cm]{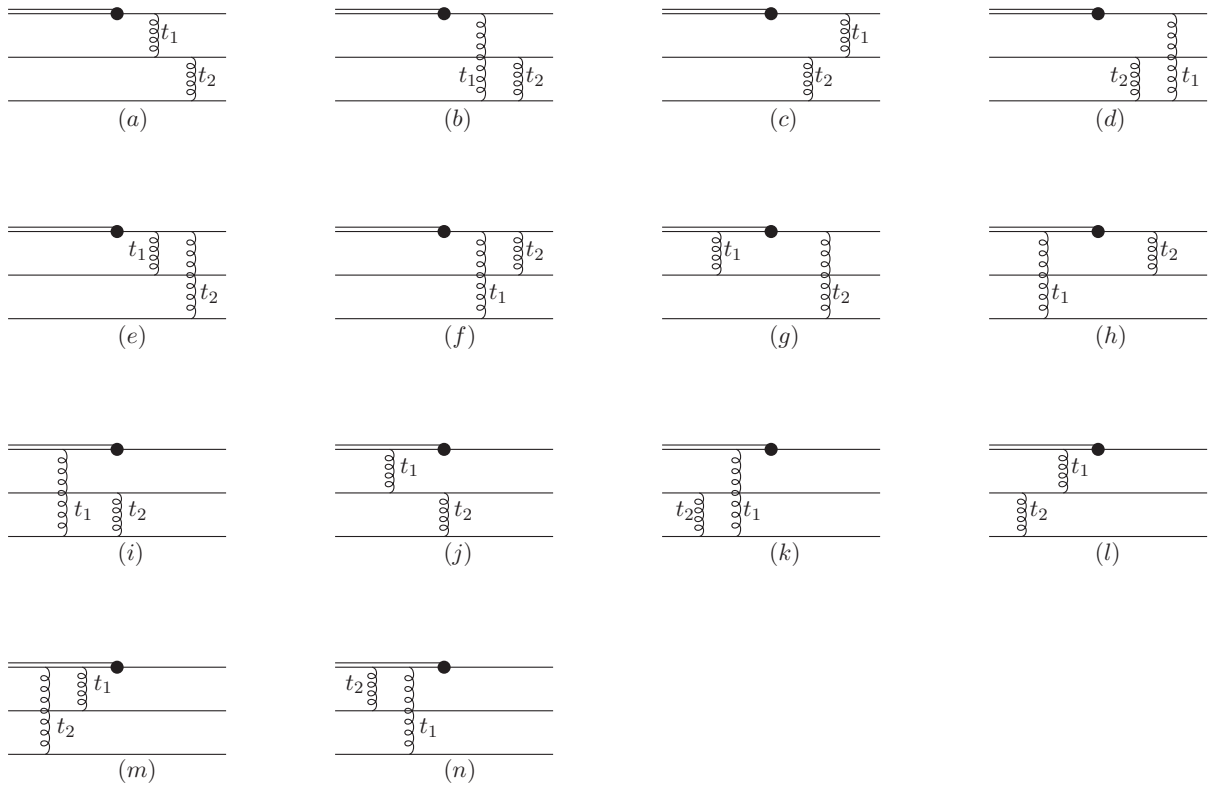}
\end{tabular}
\end{center}
\label{figure} \caption{The lowest order diagrams for the
$\Lambda_b\to \Lambda \gamma$ decay. The solid lines, double
lines, wavy lines and the black blub vertex denote the light
quarks, b quark, gluon and the electromagnetic penguin vertex,
respectively. Diagrams with triple-gluon vertex do not contribute
since their color factors are all zero in the present case.}
\label{diagrams}
\end{figure}

Collecting all contributions in Fig.\ref{diagrams} and multiplying
the corresponding Wilson coefficients, one then obtains a hard
scattering amplitude $H_{l,\mu}^{\alpha'\beta'\gamma'\alpha \beta
\gamma}(x,x',b,b',M_{\Lambda_b}) = \sum_i C_l^{eff}(t) \tilde
H_{l,\mu}^{i,\alpha'\beta'\gamma'\alpha \beta
\gamma}(x,x',b,b',M_{\Lambda_b})$. Here we have labelled the hard
scale as $t$ which is taken to be the larger of the two variables
$t_{1,2}$ associated with the virtual gluon momentum in Fig.
\ref{diagrams}, i.e. $t=\textrm{max}(t_{1}^{i},t_{2}^{i})$. The
expressions for $t_{1,2}$ are listed in Appendix C.

Finally a RG running is applied to the hard scattering amplitude
to match the scale $\nu$ in the wave functions and we
obtain\cite{Lip}
\begin{eqnarray}
&&H_{l,\mu}^{\alpha'\beta'\gamma'\alpha \beta
\gamma}(x,x',b,b',M_{\Lambda_b},\nu) =\nonumber \\
&&\textrm{exp}[-6\int_{\nu}^{t}\frac{d\overline{\mu}}
{\overline{\mu}}\gamma_q(\alpha_{s}(\overline{\mu}))]\times
H_{l,\mu}^{\alpha'\beta'\gamma'\alpha \beta
\gamma}(x,x',b,b',M_{\Lambda_b})
\end{eqnarray}

The form factors are obtained by grouping relevant terms according
to the definition in eq. (\ref{formfactor}). Using
eq.(\ref{integral}) we obtain a generic expression for the form
factors corresponding to each diagram as
\begin{eqnarray}
F^i_l&=&\sum_{j=V,A,T}\frac{\pi^2}{27}f^j_\Lambda f_{_{\Lambda_b}}
\int[Dx]\int[Db]^{i}C_l^{eff}(t^{i})\nonumber \\
&&\Psi_{\Lambda_b}(x)\Phi_{\Lambda}^{j}(x')\textrm{exp}
 [-S^{i}]H_{F}^{ij} \Omega^{i}\nonumber \\
S&=&\sum^3_{k=2}s(\omega,x_kp^-)+\sum^3_{k=1}s(\omega',x_k'p'^+)
+3\int_{\omega}^{t}\frac{d\overline{\mu}}
{\overline{\mu}}\gamma_q(\alpha_{s}(\overline{\mu}))+
3\int_{\omega'}^{t}\frac{d\overline{\mu}}
{\overline{\mu}}\gamma_q(\alpha_{s}(\overline{\mu})) \label{FF}
\end{eqnarray}
where $F^i_l$ represents the form factors contributed by the ``i''
the diagram in which operators with the Wilson coefficients
$C_l^{eff}$ are inserted, in our case $C_l^{eff} = C_7^{eff}$. The
superscript $j$ labels $V, A$, and $T$ related to the spin
structure of the valence quarks in the $\Lambda$ baryon with
$f^A_\Lambda = f^V_\Lambda = f_\Lambda$. The explicit expressions
of $\Omega^{i}$ are presented in Appendix D. The functions
$H_{F}^{ij}$ are given in Appendix E. The total form factors are
obtained by summing over contributions from all diagrams.

\section{Numerical results}

We are now ready to evaluate the form factors numerically. For
concreteness, we adopt the model proposed in
Ref.\cite{lambdabwave} for the $\Lambda_b$ baryon distribution
amplitude $\Psi$,
\begin{eqnarray}
&&\Psi(x_1,x_2,x_3) = Nx_1 x_2 x_3
\mathrm{exp}[-{M^2_{\Lambda_b}\over 2\beta^2x_1}-{m^2_q\over
2\beta^2x_2}-{m^2_q\over 2\beta^2x_3}]. \label{WPhi}
\end{eqnarray}
The normalization constant $N$ is obtained by the condition:
\begin{eqnarray}
&&\int [dx]\Psi(x_1,x_2,x_3) = 1.
\end{eqnarray}
The decay constant $f_{\Lambda_b}$ is determined by fitting
$B(\Lambda_b \to \Lambda_c l \overline{\nu})$ whose central value
is 5\% measured by DELPHI\cite{expLic} using the same PQCD method.
When fitting the data we truncate the double log Sudakov factor in
such a way that the factor exp(-s) is smaller than 1 following the
prescription in Ref.\cite{suda1}. Our numbers for $f_{\Lambda_b}$
are different from those obtained in Ref.\cite{Lic} where a
$B(\Lambda_b \to \Lambda_c l \overline{\nu})$ was taken to be 2\%.
We also have chosen cut-offs as $\omega=1.14min(1/\tilde
b_1,1/\tilde b_2,1/\tilde b_3)$ and $\omega'=1.14min(1/\tilde
b'_1,1/\tilde b'_2,1/\tilde b'_3)$. The factor 1.14 is adopted
because this cut-off choice can result in form factors which vary
smoothly with square of momentum transfer in fitting $\Lambda_b
\to \Lambda_c l \overline{\nu}$ process and it reflects the
resummation of next-to-leading double log in higher order
radiative corrections\cite{Lip2}. Also the $\beta$ and $m_q$ in
the heavy baryon wavefunction distribution need to be fixed. In
Ref.\cite{Lic,Lip,LiJ}, $\beta=1$ GeV and $m_q=0.3$ GeV were used
to estimate $\Lambda_b \to \Lambda_c l \bar{\nu}$, $\Lambda_b \to
p l \bar{\nu}$ and also $\Lambda_b \to \Lambda J/\psi$ decay
rates. $\beta$ should not be too much smaller than 1 GeV if the
form factors are dominated by perturbative contributions.
Therefore we will let both $\beta$ and $m_q$ vary within ranges as
$0.6\sim 1$GeV and $0.2\sim 0.3$GeV. The results for
$f_{\Lambda_b}$ are shown in Table \ref{flambdab} for different
parameter choices respectively.


\begin{table}[h]
\caption{Decay constant $f_{\Lambda_b}$ for different choices of
$\beta$ and $m_q$, respectively.}
\begin{center}
\begin{tabular}{|c|c|c|c|c|c|}
  \hline
  $f_{\Lambda_b}$(GeV) & $\beta=0.6$GeV & $\beta=0.7$GeV & $\beta=0.8$GeV & $\beta=0.9$GeV & $\beta=1$GeV\\
  \hline
  $m_q=0.2$GeV & $0.691\times 10^{-3}$ & $0.841\times 10^{-3}$ & $1.02\times 10^{-3}$
  & $1.21\times 10^{-3}$ & $1.43\times 10^{-3}$\\
  \hline
  $m_q=0.3$GeV & $1.27\times 10^{-3}$ & $1.45\times 10^{-3}$ & $1.65\times 10^{-3}$
  & $1.88\times 10^{-3}$ & $2.12\times 10^{-3}$\\
  \hline
\end{tabular}
\end{center}\label{flambdab}
\end{table}

The $\Lambda$ baryon distribution amplitudes have been studied
using QCD sum rules. In this work, we adopt the model proposed in
Ref.\cite{lambdawave},
\begin{eqnarray}
&&\phi^V(x_1,x_2,x_3) =
42\phi_{as}(x_1,x_2,x_3)[0.18(x_3^2-x_2^2)+0.10(x_2-x_3)],\nonumber
\\
&&\phi^A(x_1,x_2,x_3) =
-42\phi_{as}(x_1,x_2,x_3)[0.26(x_3^2+x_2^2)+0.34x_1^2-0.56x_2x_3-0.24x_1(x_2+x_3)],\nonumber
\\
&&\phi^T(x_1,x_2,x_3) =
42\phi_{as}(x_1,x_2,x_3)[1.2(x_2^2-x_3^2)-1.4(x_2-x_3)],\nonumber
\\
&&\phi_{as}(x_1,x_2,x_3) = 120x_1x_2x_3. \label{Wphi}
\end{eqnarray}
The asymmetric distribution in the momentum fractions of the three
quarks implies $SU(3)$ symmetry breaking.

The constants $f_{\Lambda}$ and $f_{\Lambda}^T$ are fixed to
be\cite{lambdawave}
\begin{eqnarray}
&&f_{\Lambda} = 0.63 \times 10^{-2}
\mathrm{GeV}^{2},\;\;f_{\Lambda}^T = 0.063 \times 10^{-2}
\mathrm{GeV}^{2}.
\end{eqnarray}

Finally to obtain the branching ratio for $\Lambda_b \to \Lambda
\gamma$, for definitiveness we fix rest of the parameters as
following. The parameter $\Lambda_{QCD}$ which enters in the
strong coupling constant and various Wilson coefficients, the b
quark mass and the CKM mixing parameters are set to be:
$\Lambda_{QCD}$ at 0.2 GeV, $m_b = 4.8 $ GeV, and the CKM mixing
parameters are set to their central values\cite{data0}: $s_{12} =
0.2243$, $s_{23} = 0.00413$, $s_{13} = 0.0037$ and $\delta_{13} =
1.05$.

Our explicit calculations show that $F_L=0$ and a non-zero value
for $F_R$ as expected since light quark and light baryon masses
have been neglected. The contributions from each diagrams for
$F_R$ are shown in Appendix E. The resulting branching ratio is
shown in Table \ref{BR}. We see that the branching ratio for
$\Lambda_b \to \Lambda \gamma$ is in the range of $(4.3\sim 6.8)
\times 10^{-8}$.

\begin{table}[h]
\caption{Branching ratio(BR) of $\Lambda_b \to \Lambda \gamma$ for
different choices of $\beta$ and $m_q$ with Sudakov truncation.}
\begin{center}
\begin{tabular}{|c|c|c|c|c|c|}
  \hline
  BR$(\times 10^8)$ & $\beta=0.6$GeV & $\beta=0.7$GeV & $\beta=0.8$GeV & $\beta=0.9$GeV & $\beta=1$GeV\\
  \hline
  $m_q=0.2$GeV & $6.76$ & $6.26$ & $6.19$ & $4.90$ & $4.67$\\
  \hline
  $m_q=0.3$GeV & $6.42$ & $5.75$ & $5.61$ & $4.44$ & $4.32$\\
  \hline
\end{tabular}
\end{center}\label{BR}
\end{table}

\section{Discussions and conclusions}

In this work we have used the perturbative QCD approach to
evaluate the branching ratio for radiative decay
$\Lambda_b\rightarrow \Lambda \gamma$. This process occurs via
penguin diagrams. Our results are shown in Table \ref{BR}. The
branching ratio obtained is much smaller than results obtained,
shown in Table \ref{ggamma}, using other methods.

\begin{table}[h]
\caption{Decay branching ratios (B) of $\Lambda_b\to \Lambda
\gamma$ based on the form factors from the QCD sum rule approach,
the covariant oscillator quark model, HQET and MIT bag model,
respectively}
\begin{center}
\begin{tabular}{|c|c|c|c|c|c|}
  \hline
  Model &pole model\cite{Mannel}& QCD sum rule\cite{Huang} & covariant oscillator quark
  model\cite{Mohanta} & HQET\cite{Cheng} & bag model\cite{Cheng} \\
  \hline
  B & $(0.10\sim 0.45)\times 10^{-5}$& $(3.7\pm 0.5)\times 10^{-5}$ &
  $0.23\times 10^{-5}$ & $(1.2\sim 1.9)\times 10^{-5}$ & $0.6\times
10^{-5}$\\
  \hline
\end{tabular}
\end{center}\label{ggamma}
\end{table}

There are uncertainties in PQCD predictions due to unknown
parameters in wavefunctions. We have tried to understand such
uncertainties by varying several relevant parameters. Within
reasonable ranges of the parameters it is not possible to obtain a
branching ratio larger than $10^{-7}$. We have considered another
possible uncertainty in the method used here. This is the choice
of the infrared cut-offs $\omega(\omega')$ in the Sudakov
suppression factor which damps the perturbative contributions. In
our calculations the cut-offs are set to the conventional values
with $\omega=1.14min(1/\tilde b_1,1/\tilde b_2,1/\tilde b_3)$ and
$\omega'=1.14min(1/\tilde b'_1,1/\tilde b'_2,1/\tilde b'_3)$
discussed in the text. The factor 1.14 is adopted because this
choice for cut-offs' can result in form factors which vary
smoothly with square of momentum transfer in fitting $\Lambda_b
\to \Lambda_c l \overline{\nu}$ process and it reflects the
resummation of next-to-leading double log in higher order
radiative corrections\cite{Lip2}. We have checked with slightly
different cut-offs and find impossible to obtain branching ratio
to be as large as what listed in Table \ref{ggamma}.

The prescription of truncating the factor exp(-s) to be smaller
than 1 described in Ref.\cite{suda1} may also be a source for
uncertainties. We therefore have evaluated the branching ratio
without this truncation. The results are shown in Table
\ref{notruncat}. We see that the results are similar to those
obtained in Table \ref{BR}.

\begin{table}[h]
\caption{Branching ratio(BR) of $\Lambda_b \to \Lambda \gamma$ for
different choices of $\beta$ and $m_q$ without Sudakov
truncation.}
\begin{center}
\begin{tabular}{|c|c|c|c|c|c|}
  \hline
  BR$(\times 10^8)$ & $\beta=0.6$GeV & $\beta=0.7$GeV & $\beta=0.8$GeV & $\beta=0.9$GeV & $\beta=1$GeV\\
  \hline
  $m_q=0.2$GeV & $8.60$ & $7.22$ & $5.91$ & $4.92$ & $4.60$\\
  \hline
  $m_q=0.3$GeV & $5.96$ & $5.73$ & $5.70$ & $4.67$ & $4.30$\\
  \hline
\end{tabular}
\end{center}\label{notruncat}
\end{table}

We therefore conclude that within the PQCD framework, the
branching ratio for $\Lambda_b \to \Lambda \gamma$ is much smaller
than other model calculations. This is somewhat surprising since
PQCD calculation for the branching ratio of $B\to K^{(*)} \gamma$
obtains a value of order consistent to other model calculations
and also agrees with experimental value of about $4\times
10^{-5}$\cite{newbgamma}. There is a huge suppression for
$\Lambda_b\to \Lambda \gamma$. At this moment there is no data
available for $\Lambda_b \to \Lambda \gamma$ yet. One has to wait
for future experimental data to tell us more. If a branching ratio
above $10^{-7}$ is measured at some future facilities, such as
LHCb, the PQCD method used here will certainly need to be
modified.

On the theoretical side, one expects the branching ratio for
$\Lambda_b\to \Lambda \gamma$ to be smaller than that of $B\to
K^{(*)}\gamma$ due to several suppression factors such as an
additional $\alpha_s^2$ and a large momentum squared $q^2$
suppression factor as one more hard gluon is exchanged between
quarks. There is also an additional Sudakov suppression factor due
to an additional spectator quark involved in the process as can be
seen from eq.(\ref{FF}).

One might question the applicability of PQCD method for the
process under consideration. One notes that in the PQCD approach,
both gluons are hard ones which excludes the possibility of
including contributions where two spectator quarks (not involved
in the weak interaction vertex) form a collective object first due
to soft gloun exchanges, i.e. the diquark, and then this object
interacts with the other quark by exchanging a hard gluon. If this
contribution turns out to be the dominant one, the branching ratio
may be substantially larger. At present there is no solid
theoretical method to treat this effect yet, we do not have a
definitive answer about this. We, however, note that estimate for
$\Lambda_b \to \Lambda J/\psi$ using the same method gives a
reasonable range compared with data\cite{LiJ}. This can be taken
as a support for the applicability of the method to $\Lambda_b$
decays. Our result for $B(\Lambda_b \to \Lambda \gamma)$
represents a reasonable estimate. The branching ratio for
$\Lambda_b\to \Lambda \gamma$ is in the range of $(4.3 \sim
8.6)\times 10^{-8}$ which is smaller than predictions using other
methods listed in Table \ref{ggamma}. We have to wait for future
experiments to provide more information.

\noindent {\bf Acknowledgements}:

This work is supported in part by NNSFC, NSC and NCTS. We would
like to thank Hsiang-nan Li and Cai-Dian L\"{u} for helpful
discussions. We also thank H.n. Li for providing us with the
program for numerical integration.
\\
\\

\noindent{\bf Appendix A: the $b$ measures}\\
The ordinary b measure is defined as
\begin{eqnarray}
&&[d\mathbf{b}]={d^2\mathbf{b}\over (2\pi)^2}
\end{eqnarray}
The explicit forms of $[D\mathbf{b}]^i$ for each diagram $i$ in
Fig.~1 are given by
\begin{eqnarray}
&&[D\mathbf{b}]^{(a)}=
[d\mathbf{b}_1][d\mathbf{b}_3][d\mathbf{b}_1'][d\mathbf{b}_3'],\;\;
[D\mathbf{b}]^{(b)}=
[d\mathbf{b}_1][d\mathbf{b}_2][d\mathbf{b}_1'][d\mathbf{b}_2'],\nonumber\\
&&[D\mathbf{b}]^{(c)}=
[d\mathbf{b}_1][d\mathbf{b}_3][d\mathbf{b}_1'][d\mathbf{b}_3'],\;\;
[D\mathbf{b}]^{(d)}=
[d\mathbf{b}_1][d\mathbf{b}_2][d\mathbf{b}_1'][d\mathbf{b}_2'],\nonumber\\
&&[D\mathbf{b}]^{(e)}=
[d\mathbf{b}_2][d\mathbf{b}_2'][d\mathbf{b}_3'],\;\;\;\;\;\;\;\;\;\;
[D\mathbf{b}]^{(f)}=
[d\mathbf{b}_3][d\mathbf{b}_2'][d\mathbf{b}_3'],\nonumber\\
&&[D\mathbf{b}]^{(g)}=
[d\mathbf{b}_2][d\mathbf{b}_3][d\mathbf{b}_3'],\;\;\;\;\;\;\;\;
\;\;[D\mathbf{b}]^{(h)}= [d\mathbf{b}_2]
[d\mathbf{b}_3][d\mathbf{b}_2']\nonumber\\
&&[D\mathbf{b}]^{(i)}=
[d\mathbf{b}_1][d\mathbf{b}_2][d\mathbf{b}_1'][d\mathbf{b}_2'],
\;\;[D\mathbf{b}]^{(j)}=
[d\mathbf{b}_1][d\mathbf{b}_3][d\mathbf{b}_1'][d\mathbf{b}_3'],\nonumber\\
&&[D\mathbf{b}]^{(k)}=
[d\mathbf{b}_1][d\mathbf{b}_2][d\mathbf{b}_1'][d\mathbf{b}_2'],\;\;
[D\mathbf{b}]^{(l)}= [d\mathbf{b}_1][d\mathbf{b}_3][d\mathbf{b}_1'][d\mathbf{b}_3'],\nonumber \\
&&[D\mathbf{b}]^{(m)}=
[d\mathbf{b}_2][d\mathbf{b}_3][d\mathbf{b}_2'],\;\;\;\;\;\;\;\;\;
[D\mathbf{b}]^{(n)}=
[d\mathbf{b}_2][d\mathbf{b}_3][d\mathbf{b}_3'].
\end{eqnarray}

\noindent{\bf Appendix B: Hard scattering amplitudes
$H_{l,\mu}^{i,\alpha'\beta'\gamma'\alpha\beta\gamma}(x_i,x_i',\textbf{k}_{T},\textbf{k}_{T}'
,M_{\Lambda_b})$}

Expressions of amplitude
$H_{l,\mu}^{i,\alpha'\beta'\gamma'\alpha\beta\gamma}(x_i,x_i',\textbf{k}_{T},\textbf{k}_{T}'
,M_{\Lambda_b})$ for each diagram in Fig. 1. In the following
$O^l_\mu$ comes from the $\gamma$-matrix in the effective
Hamiltonian, $O^{l=7}_\mu=\sigma_{\mu\nu}q^\nu R$

For the hard amplitude of Fig.1(a):
\begin{eqnarray}
&&
H_\mu^{a,\alpha'\beta'\gamma'\alpha\beta\gamma}(x_i,x_i',\textbf{k}_{T},\textbf{k}_{T}'
,M_{\Lambda_b})=\nonumber
\\
&&[\varepsilon^{abc}\varepsilon^{a'b'c'}(T^j)_{c'c}(T^jT^i)_{b'b}(T^i)_{a'a}]g_s^4
{(\gamma_\rho)_{\gamma'\gamma}[\gamma^\rho(\rlap /p'-\rlap
/k'_1-\rlap
/k_3)\gamma^\lambda]_{\beta'\beta}[\gamma_\lambda(\rlap /p'-\rlap
/p+\rlap /k_1)O_\mu]_{\alpha'\alpha}\over
(p'-k_1'-k_3)^2(p'-p+k_1)^2(p-p'+k_1'-k_1)^2(k_3-k_3')^2}\nonumber
\\
&&=C_N g_s^4 {(\gamma_\rho)_{\gamma'\gamma}[\gamma^\rho(\rlap
/p'-\rlap /k'_1-\rlap
/k_3)\gamma^\lambda]_{\beta'\beta}[\gamma_\lambda(\rlap /p'-\rlap
/p+\rlap /k_1)O_\mu]_{\alpha'\alpha}\over
[A_a+(\textbf{k}_{1T}'+\textbf{k}_{3T})^2][B_a+\textbf{k}_{1T}^2][C_a+(\textbf{k}_{1T}
-\textbf{k}_{1T}')^2][D_a+(\textbf{k}_{3T}-\textbf{k}_{3T}')^2]}
\end{eqnarray}
with
\begin{eqnarray}
&&A_a=x_3(1-x_1')\rho M_{\Lambda_b}^2, B_a=(1-x_1)\rho
M_{\Lambda_b}^2, C_a=(1-x_1)(1-x_1')\rho M_{\Lambda_b}^2,
D_a=x_3x_3'\rho M_{\Lambda_b}^2
\end{eqnarray}
and the color factor
\begin{eqnarray}
&&C_N=\varepsilon^{abc}\varepsilon^{a'b'c'}(T^j)_{c'c}(T^jT^i)_{b'b}(T^i)_{a'a}={(N^2-1)(N+1)\over
12}
\end{eqnarray} For the hard amplitude of Fig.1(b):
\begin{eqnarray}
&&H_\mu^{b,\alpha'\beta'\gamma'\alpha\beta\gamma}(x_i,x_i',\textbf{k}_{T},\textbf{k}_{T}'
,M_{\Lambda_b})=\nonumber
\\
&&[\varepsilon^{abc}\varepsilon^{a'b'c'}(T^iT^j)_{c'c}(T^i)_{b'b}(T^j)_{a'a}]g_s^4
{(\gamma^\rho)_{\beta'\beta}[\gamma_\rho(\rlap /p'-\rlap
/k_1'-\rlap
/k_2)\gamma_\lambda]_{\gamma'\gamma}[\gamma^\lambda(\rlap
/p'-\rlap /p+\rlap /k_1)O_\mu]_{\alpha'\alpha}\over
(p'-k_1'-k_2)^2(p'-p+k_1)^2(p-p'-k_1+k_1')^2(k_2'-k_2)^2}\nonumber
\\
&&=C_Ng_s^4 {(\gamma^\rho)_{\beta'\beta}[\gamma_\rho(\rlap
/p'-\rlap /k_1'-\rlap
/k_2)\gamma_\lambda]_{\gamma'\gamma}[\gamma^\lambda(\rlap
/p'-\rlap /p+\rlap /k_1)O_\mu]_{\alpha'\alpha}\over
[A_b+(\textbf{k}_{1T}'+\textbf{k}_{2T})^2][B_b+\textbf{k}_{1T}^2][C_b+(\textbf{k}_{1T}
-\textbf{k}_{1T}')^2][D_b+(\textbf{k}_{2T}-\textbf{k}_{2T}')^2]}
\end{eqnarray}
with
\begin{eqnarray}
&&A_b=x_2(1-x_1')\rho M_{\Lambda_b}^2, B_b=(1-x_1)\rho
M_{\Lambda_b}^2, C_b=(1-x_1)(1-x_1')\rho M_{\Lambda_b}^2,
D_b=x_2x_2'\rho M_{\Lambda_b}^2
\end{eqnarray}

Inspection of the above calculations, one notices that one can
easily obtain
$H_\mu^{b,\alpha'\beta'\gamma'\alpha\beta\gamma}(x_i,x_i',\textbf{k}_{T},\textbf{k}_{T}'
,M_{\Lambda_b})$ from
$H_\mu^{a,\alpha'\beta'\gamma'\alpha\beta\gamma}(x_i,x_i',\textbf{k}_{T},\textbf{k}_{T}'
,M_{\Lambda_b})$ and vice versa by simply exchanging the momentum
indices 2 and 3 for $\textbf{k}$ and $\textbf{k}'$, and exchanging
the positions of the Dirac indices $\gamma' \gamma$ and
$\beta'\beta$. Due to these properties, the contributions to the
form factors from the above two diagrams are the same. This fact
can be easily understood by noticing the following properties of
the quantities related to the distribution amplitudes: i) The
distribution amplitudes $\Psi(x_1,x_2,x_3)$, and
$\phi^{A}(x_1,x_2,x_3)$ are symmetric in exchanging $x_2$ and
$x_3$, while $\phi^{V,T}(x_1,x_2,x_3)$ are anti-symmetric in
exchanging $x_2$ and $x_3$, as can be seen from eqs.(\ref{WPhi})
and (\ref{Wphi}). And ii) When exchanging the Dirac indices
$\beta$ and $\gamma$, the expressions for
$(Y_{\Lambda_b})_{\alpha\beta\gamma}(k_i\nu)$ in eq.(\ref{YYb}),
and terms proportional to $\phi^A$ for
$(Y_\Lambda)_{\alpha\beta\gamma}(k'_i,\nu)$ in eq.(\ref{YY}) will
have a sign change, while terms proportional to $\phi^{V,T}$
remain the same. Since going from the contribution of diagram (a)
to diagram (b) involves both actions: exchanging the momentum
indices 2 and 3, and the Dirac indices $\beta$ and $\gamma$, this
results in no sign changes for all the terms involved. After
integrating out $x(x')_{2,3}$ and $b(b')_{2,3}$ to obtain the
final form factors using eq.(\ref{FF}), one then obtains the same
results for both diagrams (a) and (b).

Similar situation happens for the following pairs of diagrams: (c)
and (d), (e) and (f), (g) and (h), (i) and (j), (k) and (l), and,
(m) and (n).  In the following we will only display the results
for diagrams (a), (c), (e), (g), (i), (k) and (m). The expressions
for diagrams (b), (d), (f), (h), (j), (l) and (n) can be obtained
by exchanging $x(x')_2$ and $x(x')_3$, and also $\gamma'\gamma$
and $\beta' \beta$.

For the hard amplitude of Fig.1(c):
\begin{eqnarray}
&&H_\mu^{c,\alpha'\beta'\gamma'\alpha\beta\gamma}(x_i,x_i',\textbf{k}_{T},\textbf{k}_{T}'
,M_{\Lambda_b})=\nonumber
\\
&&[\varepsilon^{abc}\varepsilon^{a'b'c'}(T^j)_{c'c}(T^iT^j)_{b'b}(T^i)_{a'a}]g_s^4
{(\gamma_\rho)_{\gamma'\gamma}[\gamma^\lambda(\rlap /p-\rlap
/k_1-\rlap /k_3')\gamma^\rho]_{\beta'\beta}[\gamma_\lambda(\rlap
/p'-\rlap /p+\rlap /k_1)O_\mu]_{\alpha'\alpha}\over
(p-k_1-k_3')^2(p'-p+k_1)^2(p-p'+k_1'-k_1)^2(k_3-k_3')^2}\nonumber
\\
&&=C_Ng_s^4 {(\gamma_\rho)_{\gamma'\gamma}[\gamma^\lambda(\rlap
/p-\rlap /k_1-\rlap
/k_3')\gamma^\rho]_{\beta'\beta}[\gamma_\lambda(\rlap /p'-\rlap
/p+\rlap /k_1)O_\mu]_{\alpha'\alpha}\over
[A_c+(\textbf{k}_{1T}+\textbf{k}_{3T}')^2][B_c+\textbf{k}_{1T}^2][C_c+(\textbf{k}_{3T}-\textbf{k}_{3T}')^2]
[D_c+(\textbf{k}_{1T} -\textbf{k}_{1T}')^2]}
\end{eqnarray}
with
\begin{eqnarray}
&&A_c=x_3'(1-x_1)\rho M_{\Lambda_b}^2, B_c=(1-x_1)\rho
M_{\Lambda_b}^2, C_c=x_3x_3'\rho M_{\Lambda_b}^2,
D_c=(1-x_1)(1-x_1')\rho M_{\Lambda_b}^2
\end{eqnarray}

For the hard amplitude of Fig.1(e):
\begin{eqnarray}
&&H_\mu^{e,\alpha'\beta'\gamma'\alpha\beta\gamma}(x_i,x_i',\textbf{k}_{T},\textbf{k}_{T}'
,M_{\Lambda_b})=\nonumber
\\
&&[\varepsilon^{abc}\varepsilon^{a'b'c'}(T^i)_{c'c}(T^iT^j)_{b'b}(T^j)_{a'a}]g_s^4
{(\gamma_\rho)_{\gamma'\gamma}[\gamma^\rho(\rlap /p'-\rlap
/k_2'-\rlap /k_3)\gamma^\lambda(\rlap /p'-\rlap /p+\rlap
/k_1)O_\mu]_{\alpha'\alpha}(\gamma_\lambda)_{\beta'\beta}\over
(p'-k_2'-k_3)^2(p'-p+k_1)^2(k_2'-k_2)^2(k_3-k_3')^2}\nonumber
\\
&&=C_Ng_s^4 {(\gamma_\rho)_{\gamma'\gamma}[\gamma^\rho(\rlap
/p'-\rlap /k_2'-\rlap /k_3)\gamma^\lambda(\rlap /p'-\rlap /p+\rlap
/k_1)O_\mu]_{\alpha'\alpha}(\gamma_\lambda)_{\beta'\beta}\over
[A_e+(\textbf{k}_{2T}'+\textbf{k}_{3T})^2][B_e+\textbf{k}_{1T}^2][C_e+(\textbf{k}_{2T}
-\textbf{k}_{2T}')^2][D_e+(\textbf{k}_{3T}-\textbf{k}_{3T}')^2]}
\end{eqnarray}
with
\begin{eqnarray}
&&A_e=x_3(1-x_1')\rho M_{\Lambda_b}^2, B_e=(1-x_1)\rho
M_{\Lambda_b}^2, C_e=x_2x_2'\rho M_{\Lambda_b}^2, D_e=x_3x_3'\rho
M_{\Lambda_b}^2
\end{eqnarray}

For the hard amplitude of Fig.1(g):
\begin{eqnarray}
&&H_\mu^{g,\alpha'\beta'\gamma'\alpha\beta\gamma}(x_i,x_i',\textbf{k}_{T},\textbf{k}_{T}'
,M_{\Lambda_b})=\nonumber
\\
&&[\varepsilon^{abc}\varepsilon^{a'b'c'}(T^i)_{c'c}(T^j)_{b'b}(T^iT^j)_{a'a}]g_s^4
{(\gamma_\rho)_{\gamma'\gamma}[\gamma^\rho(\rlap /p'-\rlap
/k_2'-\rlap /k_3)O_\mu(\rlap /p-\rlap /k_3-\rlap
/k_2'+m_b)\gamma^\lambda]_{\alpha'\alpha}(\gamma_\lambda)_{\beta'\beta}\over
[(p-k_3-k_2')^2-m_b^2](p'-k_2'-k_3)^2(k_2'-k_2)^2(k_3'-k_3)^2}\nonumber
\\
&&=C_Ng_s^4 {(\gamma_\rho)_{\gamma'\gamma}[\gamma^\rho(\rlap
/p'-\rlap /k_2'-\rlap /k_3)O_\mu(\rlap /p-\rlap /k_3-\rlap
/k_2'+m_b)\gamma^\lambda]_{\alpha'\alpha}(\gamma_\lambda)_{\beta'\beta}\over
[A_g+(\textbf{k}_{2T}'+\textbf{k}_{3T})^2][B_g+(\textbf{k}_{2T}'+\textbf{k}_{3T})^2]
[C_g+(\textbf{k}_{2T}-\textbf{k}_{2T}')^2][D_g+(\textbf{k}_{3T}-\textbf{k}_{3T}')^2]}
\end{eqnarray}
with
\begin{eqnarray}
&&A_g=(x_2'(1-x_3)\rho+x_3) M_{\Lambda_b}^2, B_g=x_3(1-x_2')\rho
M_{\Lambda_b}^2, C_g=x_2x_2'\rho M_{\Lambda_b}^2, D_g=x_3x_3'\rho
M_{\Lambda_b}^2
\end{eqnarray}

For the hard amplitude of Fig.1(i):
\begin{eqnarray}
&&H_\mu^{i,\alpha'\beta'\gamma'\alpha\beta\gamma}(x_i,x_i',\textbf{k}_{T},\textbf{k}_{T}'
,M_{\Lambda_b})=\nonumber
\\
&&[\varepsilon^{abc}\varepsilon^{a'b'c'}(T^iT^j)_{c'c}(T^i)_{b'b}(T^j)_{a'a}]g_s^4
{[\gamma_\rho(\rlap /p'-\rlap /k_1'-\rlap
/k_2)\gamma_\lambda]_{\gamma'\gamma}[O_\mu(\rlap /p-\rlap
/p'-\rlap
/k_1'+m_b)\gamma^\lambda]_{\alpha'\alpha}(\gamma^\rho)_{\beta'\beta}\over
[(p-p'+k_1')^2-m_b^2](p'-k_1'-k_2)^2(p-p'-k_1+k_1')^2(k_2'-k_2)^2}\nonumber
\\
&&=C_Ng_s^4 {[\gamma_\rho(\rlap /p'-\rlap /k_1'-\rlap
/k_2)\gamma_\lambda]_{\gamma'\gamma}[O_\mu(\rlap /p-\rlap
/p'-\rlap
/k_1'+m_b)\gamma^\lambda]_{\alpha'\alpha}(\gamma^\rho)_{\beta'\beta}\over
[A_i+(\textbf{k}_{2T}+\textbf{k}_{1T}')^2][B_i+(\textbf{k}_{1T}'+\textbf{k}_{2T})^2][C_i
+(\textbf{k}_{1T}-\textbf{k}_{1T}')^2][D_i+(\textbf{k}_{2T}-\textbf{k}_{2T}')^2]}
\end{eqnarray}
with
\begin{eqnarray}
&&A_i=(1-x_1')\rho M_{\Lambda_b}^2, B_i=x_2(1-x_1')\rho
M_{\Lambda_b}^2, C_i=(1-x_1)(1-x_1')\rho M_{\Lambda_b}^2,
D_i=x_2x_2'\rho M_{\Lambda_b}^2
\end{eqnarray}

For the hard amplitude of Fig.1(k):
\begin{eqnarray}
&&H_\mu^{k,\alpha'\beta'\gamma'\alpha\beta\gamma}(x_i,x_i',\textbf{k}_{T},\textbf{k}_{T}'
,M_{\Lambda_b})=\nonumber
\\
&&[\varepsilon^{abc}\varepsilon^{a'b'c'}(T^iT^j)_{c'c}(T^j)_{b'b}(T^i)_{a'a}]g_s^4
{[\gamma_\rho(\rlap /p-\rlap /k_1-\rlap
/k_2')\gamma_\lambda]_{\gamma'\gamma}[O_\mu(\rlap /p-\rlap
/p'+\rlap
/k_1'+m_b)\gamma^\rho]_{\alpha'\alpha}(\gamma^\lambda)_{\beta'\beta}\over
[(p-p'+k_1')^2-m_b^2](p-k_1-k_2')^2(p-p'-k_1+k_1')^2(k_2'-k_2)^2}\nonumber
\\
&&=C_Ng_s^4 {[\gamma_\rho(\rlap /p-\rlap /k_1-\rlap
/k_2')\gamma_\lambda]_{\gamma'\gamma}[O_\mu(\rlap /p-\rlap
/p'+\rlap
/k_1'+m_b)\gamma^\rho]_{\alpha'\alpha}(\gamma^\lambda)_{\beta'\beta}\over
[A_k+\textbf{k}_{1T}'^2][B_k+(\textbf{k}_{1T}+\textbf{k}_{2T}')^2][C_k+(\textbf{k}_{1T}
-\textbf{k}_{1T}')^2][D_k+(\textbf{k}_{2T}-\textbf{k}_{2T}')^2]}
\end{eqnarray}
with
\begin{eqnarray}
&&A_k=(1-x_1')\rho M_{\Lambda_b}^2, B_k=x_2'(1-x_1)\rho
M_{\Lambda_b}^2, C_k=(1-x_1)(1-x_1')\rho M_{\Lambda_b}^2,
D_k=x_2x_2'\rho M_{\Lambda_b}^2
\end{eqnarray}

For the hard amplitude of Fig.1(m):
\begin{eqnarray}
&&H_\mu^{m,\alpha'\beta'\gamma'\alpha\beta\gamma}(x_i,x_i',\textbf{k}_{T},\textbf{k}_{T}'
,M_{\Lambda_b})=\nonumber
\\
&&[\varepsilon^{abc}\varepsilon^{a'b'c'}(T^j)_{c'c}(T^i)_{b'b}(T^iT^j)_{a'a}]g_s^4
{(\gamma_\lambda)_{\gamma'\gamma}[O_\mu(\rlap /p-\rlap /p'+\rlap
/k_1'+m_b)\gamma_\rho(\rlap /p-\rlap /k_2-\rlap
/k_3'+m_b)\gamma^\lambda]_{\alpha'\alpha}(\gamma^\rho)_{\beta'\beta}\over
[(p-p'+k_1')^2-m_b^2][(p-k_2-k_3')^2-m_b^2](k_2'-k_2)^2(k_3'-k_3)^2}\nonumber
\\
&&=C_Ng_s^4 {(\gamma_\lambda)_{\gamma'\gamma}[O_\mu(\rlap /p-\rlap
/p'+\rlap /k_1'+m_b)\gamma_\rho(\rlap /p-\rlap /k_2-\rlap
/k_3'+m_b)\gamma^\lambda]_{\alpha'\alpha}(\gamma^\rho)_{\beta'\beta}\over
[A_m+\textbf{k}_{1T}'^2][B_m+(\textbf{k}_{2T}+\textbf{k}_{3T}')^2][C_m+(\textbf{k}_{2T}
-\textbf{k}_{2T}')^2][D_m+(\textbf{k}_{3T}-\textbf{k}_{3T}')^2]}
\end{eqnarray}
with
\begin{eqnarray}
&&A_m=(1-x_1')\rho M_{\Lambda_b}^2, B_m=(x_3'(1-x_2)\rho+x_2)
M_{\Lambda_b}^2, C_m=x_2x_2'\rho M_{\Lambda_b}^2, D_m=x_3x_3'\rho
M_{\Lambda_b}^2
\end{eqnarray}

The expressions for the hard scattering amplitude in $b$ and $b'$
space are obtained by making a Fourier transformation on $k_T$ and
$k'_T$ space. In the following we given one example for Fig.1(a)
as an illustration. We note that the $k_T$ and $k'_T$ dependencies
are all in the denominators in the above expressions, one then
just needs to consider that part of the fourier transformation.
For Fig.1(a), it is given by
\begin{eqnarray}
&&\Omega^{(a)}(x_i,x_i',\textbf{k}_{T},\textbf{k}_{T}',M_{\Lambda_b})
={1\over
[A_a+(\textbf{k}_{1T}'+\textbf{k}_{3T})^2][B_a+\textbf{k}_{1T}^2][C_a+(\textbf{k}_{1T}-\textbf{k}_{1T}')^2]
[D_a+(\textbf{k}_{3T}-\textbf{k}_{3T}')^2]}.
\end{eqnarray}

The fourier transformed expression is then given by
\begin{eqnarray}
\Omega^{(a)}(x_i,x_i',b_i,b_i',M_{\Lambda_b}) = \int
e^{-i(\textbf{k}_{1T}\cdot \textbf{b}_{1}+\textbf{k}_{1T}'\cdot
\textbf{b}_{1}'+\textbf{k}_{3T}\cdot
\textbf{b}_{3}+\textbf{k}_{3T}'\cdot
\textbf{b}_{3}')}\Omega^{(a)}(x_i,x_i',\textbf{k}_{T},\textbf{k}_{T}',M_{\Lambda_b})
d^2\textbf{k}_{1T}d^2\textbf{k}_{1T}'d^2\textbf{k}_{3T}d^2\textbf{k}_{3T}'.
\end{eqnarray}

Defining $\textbf{k}_{AT}\equiv\textbf{k}_{1T}'+\textbf{k}_{3T}$,
$\textbf{k}_{BT}\equiv\textbf{k}_{1T}$,
$\textbf{k}_{CT}\equiv\textbf{k}_{1T}-\textbf{k}_{1T}'$,
 and $\textbf{k}_{DT}\equiv\textbf{k}_{3T}-\textbf{k}_{3T}'$, we
 rewrite the transformation as
\begin{eqnarray}
&&\Omega^{(a)}(x_i,x_i',b_i,b_i',M_{\Lambda_b})\nonumber\\
&&=\int {e^{-i[\textbf{k}_{AT}\cdot
(\textbf{b}_{3}+\textbf{b}_{3}')+\textbf{k}_{BT}\cdot
(\textbf{b}_{1}+\textbf{b}_{1}'-\textbf{b}_{3}-\textbf{b}_{3}')+\textbf{k}_{CT}\cdot
(-\textbf{b}_{1}'+\textbf{b}_{3}+\textbf{b}_{3}')+\textbf{k}_{DT}\cdot
(-\textbf{b}_{3}')]} 1\over
(\textbf{k}_{AT}^2+A_a)(\textbf{k}_{BT}^2+B_a)(\textbf{k}_{CT}^2+C_a)(\textbf{k}_{DT}^2+D_a)}
d^2\textbf{k}_{AT}d^2\textbf{k}_{BT}d^2\textbf{k}_{CT}d^2\textbf{k}_{DT}\nonumber
\\
&&=(2\pi)^4K_0(\sqrt{A_a}|b_3+b_3'|)K_0(\sqrt{B_a}|b_1+b_1'-b_3-b_3'|)
K_0(\sqrt{C_a}|b_1'-b_3-b_3'|)K_0(\sqrt{D_a}|b_3'|),
\end{eqnarray}

In the above we have used
\begin{eqnarray}
&&\int d^2k{e^{i\mathbf{k}\cdot \mathbf{b}}\over k^2+A}=2\pi
K_0(\sqrt{A}|\mathbf{b}|), A>0.
\end{eqnarray}

One obtains the expression for
$H_\mu^{a,\alpha'\beta'\gamma'\alpha\beta\gamma}(x_i,x_i',b,b'
,M_{\Lambda_b})$ as
\begin{eqnarray}
&&H_\mu^{a,\alpha'\beta'\gamma'\alpha\beta\gamma}(x_i,x_i',b, b'
,M_{\Lambda_b})=C_Ng_s^4
{(\gamma_\rho)_{\gamma'\gamma}[\gamma^\rho(\rlap /p'-\rlap
/k'_1-\rlap
/k_3)\gamma^\lambda]_{\beta'\beta}[\gamma_\lambda(\rlap /p'-\rlap
/p+\rlap /k_1)O_\mu]_{\alpha'\alpha}
\widetilde{\Omega^{(a)}}(x_i,x_i',b_i,b_i',M_{\Lambda_b})}.\nonumber\\
\end{eqnarray}

In carrying out the fourier transformations for other diagrams,
two other forms of functions will be encountered. We list them in
the following
\begin{eqnarray}
&&\int d^2k{e^{i\mathbf{k}\cdot \mathbf{b}}\over
(k^2+A)(k^2+B)}=\pi\int^1_0dz{|\mathbf{b}|K_1(\sqrt{Z_1}|\mathbf{b}|)\over
\sqrt{Z_1}}, A, B>0, \nonumber \\
&&\int d^2k_1 d^2k_2{e^{i(\mathbf{k_1}\cdot
\mathbf{b_1}+\mathbf{k_2}\cdot \mathbf{b_2})}\over
(k_1^2+A)(k_2^2+B)[(k_1+k_2)^2+C]}=\pi^2 \int^1_0{dz_1 dz_2\over
z_1(1-z_1)}{\sqrt{X_2}\over \sqrt{|Z_2|}}K_1(\sqrt{X_2Z_2}),
\end{eqnarray}
where $A>0$ and $B$, $C$ arbitrary. $K_0$ and $K_1$ are the
modified Bessel functions of the second kind. And
\begin{eqnarray}
&&Z_1=Az+B(1-z), \nonumber \\
&&Z_2=A(1-z_2)+{z_2\over z_1(1-z_1)}[B(1-z_1)+Cz_1],\nonumber \\
&&X_2=(\mathbf{b}_1-z_1\mathbf{b}_2)^2+{z_1(1-z_1)\over
z_2}\mathbf{b}_2^2
\end{eqnarray}

\noindent{\bf Appendix C: The maximum of $t_{1,2}$}

The hard scales, the maximal of $t_1^i$ and $t_2^i$ for diagrams
(a), (c), (e), (g), (i), (k), and (m) in Fig.~1. Exchanging
$b(b')_2$ and $b(b')_3$, one obtains the expressions for diagrams
(b), (d), (f), (h), (j), (l) and (n). The expressions of $C_i,
D_i$ are collected in Appendix B.

\begin{table}[h]
\begin{center}
\renewcommand{\arraystretch}{0.5}
\tabcolsep=0.05cm
\begin{tabular}{|c|c|c|}
  \hline
  i & $t_1^i$ & $t_2^i$ \\
  \hline
  (a) & $\textrm{max}\{\sqrt{|C_a|},\frac{1}{|\mathbf{b}_1'-\mathbf{b}_3-\mathbf{b}_3'|},\omega,\omega'\}$ &
  $\textrm{max}\{\sqrt{|D_a|},\frac{1}{|\mathbf{b}_3'|},\omega,\omega'\}$\\
  (c) & $\textrm{max}\{\sqrt{|C_c|},\frac{1}{|\mathbf{b}_1'|},\omega,\omega'\}$ &
  $\textrm{max}\{\sqrt{|D_c|},\frac{1}{|\mathbf{b}_3|},\omega,\omega'\}$  \\
  (e) & $\textrm{max}\{\sqrt{|C_e|},\frac{1}{|\mathbf{b}_2'|},\omega,\omega'\}$ &
   $\textrm{max}\{\sqrt{|D_e|},\frac{1}{|\mathbf{b}_3'|},\omega,\omega'\}$\\
  (g) & $\textrm{max}\{\sqrt{|C_g|},\frac{1}{|\mathbf{b}_2|},\omega,\omega'\}$ &
  $\textrm{max}\{\sqrt{|D_g|},\frac{1}{|\mathbf{b}_3'|},\omega,\omega'\}$ \\
  (i) & $\textrm{max}\{\sqrt{|C_i|},\frac{1}{|\mathbf{b}_1|},\omega,\omega'\}$ &
  $\textrm{max}\{\sqrt{|D_i|},\frac{1}{|\mathbf{b}_2'|},\omega,\omega'\}$\\
  (k) & $\textrm{max}\{\sqrt{|C_k|},\frac{1}{|\mathbf{b}_1-\mathbf{b}_2-\mathbf{b}_2'|},\omega,\omega'\}$ &
  $\textrm{max}\{\sqrt{|D_k|},\frac{1}{|\mathbf{b}_2|},\omega,\omega'\}$\\
  (m) & $\textrm{max}\{\sqrt{|C_m|},\frac{1}{|\mathbf{b}_2'|},\omega,\omega'\}$ &
   $\textrm{max}\{\sqrt{|D_m|},\frac{1}{|\mathbf{b}_3|},\omega,\omega'\}$\\
  \hline
 \end{tabular}
\end{center}
\end{table}

\noindent{\bf Appendix D: Expressions of $\Omega^i$}\\
The  expression of $\Omega^i$ for diagrams (a), (c), (e), (g),
(i), (k), and (m) in Fig.~1. Exchanging $b(b')_2$ and $b(b')_3$,
one obtains the expressions for diagrams (b), (d), (f), (h), (j),
(l) and (n).
\begin{eqnarray}
&&\Omega^{(a)} =
(2\pi)^4K_0(\sqrt{A_a}|\mathbf{b}_3+\mathbf{b}_3'|)K_0(\sqrt{B_a}|\mathbf{b}_1
+\mathbf{b}_1'-\mathbf{b}_3-\mathbf{b}_3'|)
K_0(\sqrt{C_a}|\mathbf{b}_1'-\mathbf{b}_3-\mathbf{b}_3'|)
K_0(\sqrt{D_a}|\mathbf{b}_3'|)\nonumber \\
&&\Omega^{(c)}
=(2\pi)^4K_0(\sqrt{A_c}|\mathbf{b}_3+\mathbf{b}_3'|)K_0(\sqrt{B_c}|\mathbf{b}_1
+\mathbf{b}_1'+\mathbf{b}_3+\mathbf{b}_3'|)
K_0(\sqrt{C_c}|\mathbf{b}_1'|)K_0(\sqrt{D_c}|\mathbf{b}_3|)\nonumber \\
&&\Omega^{(e)} = 8\pi^5\int^1_0 dz_1dz_2 {1\over
z_1(1-z_1)}{\sqrt{X_2^e}\over
\sqrt{|Z_2^e|}}K_1(\sqrt{X_2^eZ_2^e})K_0(\sqrt{D_e}|\mathbf{b}_3'|)\nonumber \\
&&\Omega^{(g)} = 16\pi^5\int^1_0
dz{|\mathbf{b}_3+\mathbf{b}_3'|K_1(\sqrt{Z_1^g}|\mathbf{b}_3+\mathbf{b}_3'|)\over
\sqrt{Z_1^g}}
K_0(\sqrt{C_g}|\mathbf{b}_2|)K_0(\sqrt{D_g}|\mathbf{b}_3'|)\nonumber \\
&&\Omega^{(i)} =
(2\pi)^4K_0(\sqrt{A_i}|\mathbf{b}_1+\mathbf{b}_1'-\mathbf{b}_2-\mathbf{b}_2'|)
K_0(\sqrt{B_i}|\mathbf{b}_2+\mathbf{b}_2'|)
K_0(\sqrt{C_i}|\mathbf{b}_1|)K_0(\sqrt{D_i}|\mathbf{b}_2'|)\nonumber \\
&&\Omega^{(k)}
=(2\pi)^4K_0(\sqrt{A_k}|\mathbf{b}_1+\mathbf{b}_1'-\mathbf{b}_2-\mathbf{b}_2'|)
K_0(\sqrt{B_k}|\mathbf{b}_2+\mathbf{b}_2'|)
K_0(\sqrt{C_k}|\mathbf{b}_1-\mathbf{b}_2-\mathbf{b}_2'|)
K_0(\sqrt{D_k}|\mathbf{b}_2|)\nonumber \\
&&\Omega^{(m)} = 8\pi^5\int^1_0 dz_1dz_2{1\over
z_1(1-z_1)}{\sqrt{X_2^m}\over
\sqrt{|Z_2^m|}}K_1(\sqrt{X_2^mZ_2^m})K_0(\sqrt{D_m}|\mathbf{b}_3|)\nonumber \\
\end{eqnarray}
with
\begin{eqnarray}
&&X_2^e=(\mathbf{b}_2'+z_1\mathbf{b}_2)^2+{z_1(1-z_1)\over
z_2}\mathbf{b}_2^2, Z_2^e=A_e(1-z_2)+{z_2\over
z_1(1-z_1)}[B_e(1-z_1)+C_ez_1]\nonumber
\\
&&Z_1^g=A_gz+B_g(1-z)\nonumber \\
&&X_2^m=(\mathbf{b}_2'+z_1\mathbf{b}_2)^2+{z_1(1-z_1)\over
z_2}\mathbf{b}_2^2, Z_2^m=A_m(1-z_2)+{z_2\over z_1(1-z_1)}
[B_m(1-z_1)+C_mz_1]\nonumber \\
\end{eqnarray}

\noindent{\bf Appendix E: Expressions for $H_F^{ij}$}\\

In this appendix we list $H^{ij}_F$ corresponding to the form
factors defined in eq.(\ref{FF}). We use $\tilde F_R^j$ for each
diagram. The expressions for diagrams (a), (e), (g), (i), (k), and
(m) in Fig.~1, whenever non-zero, are listed in the following. The
expressions for diagrams (b), (f), (h), (j), (l) and (n) can be
obtained by exchanging $x(x')_2$ and $x(x')_3$ and changing the
signs for expressions $F_R^{V,T}$. Diagrams (c) and (d) have no
contributions to $\Lambda_b\to \Lambda \gamma$. $\tilde F_L$ is
equal to zero in our approximation.

For the hard amplitudes of Fig.1(a):
\begin{eqnarray}
&&\widetilde{F_R^A}=8x_3\rho^2M_{\Lambda_b}^4
\end{eqnarray}

The relation between the tilde form factors listed above and the
form factors in eq.(\ref{formfactor}) is as the following, taking
$\tilde F_R^A$ as an example, $F^A_R =\frac{\pi^2}{27}f^j_\Lambda
f_{_{\Lambda_b}} \int[Dx]\int[Db]^{i}C_l^{eff}(t^{i})
\Psi_{\Lambda_b}(x)\Phi_{\Lambda}^{j}(x')\textrm{exp}
 [-S^{i}]\tilde F^A_R \Omega^{i}$. For this example $j = A$, and $f^A_\Lambda = f_\Lambda$.
 For Fig.1(a), `i' takes the
 value `a'. Similar for other form factors and diagrams.

The other non-zero contributions are
\begin{eqnarray}
Fig.1(e): &&\widetilde{F_R^V}=
\widetilde{F_R^A}=-4M_{\Lambda_b}^4\rho^2x_3,\nonumber\\
Fig.1(g):
&&\widetilde{F_R^V}=4M_{\Lambda_b}^3(m_b\rho+M_{\Lambda_b}(-2x_3(-1+\rho)+(1+x_2'(-1+\rho))\rho))\nonumber \\
&&\widetilde{F_R^A}=4M_{\Lambda_b}^3(-1+\rho)(m_b+M_{\Lambda_b}(-1+x_2'\rho)),\nonumber\\
Fig.1(i):
&&\widetilde{F_R^A}=8M_{\Lambda_b}^3x_2(m_b(-1+\rho)+M_{\Lambda_b}(1+(-1+x_1')\rho)),\nonumber\\
Fig.1(k): &&\widetilde{F_R^A}=8M_{\Lambda_b}^3\rho
x_2'(m_b+M_{\Lambda_b} (-1+\rho)),\nonumber\\
Fig.1(m):
&&\widetilde{F_R^V}=-4M_{\Lambda_b}^2(-m_b^2(-2+\rho)+M_{\Lambda_b}m_b(x_2-
x_2\rho+(-1+x_1'+x_3')\rho)\nonumber \\
&&\;\;\;\;\;\;+M_{\Lambda_b}^2(-2+(2-x_1'+x_3')\rho-x_3'\rho^2+x_2(1+(-1+x_1')\rho)))\nonumber \\
&&\widetilde{F_R^A}=-4M_{\Lambda_b}^2(m_b^2\rho+M_{\Lambda_b}m_b(-2+
x_2(1+\rho)+(-1+x_1'+x_3')\rho)\nonumber
\\
&&\;\;\;\;\;\;+M_{\Lambda_b}^2(2-(2+x_1'+x_3')\rho+
x_3'\rho^2+x_2(-1+(1+x_1')\rho)))
\end{eqnarray}

\end{document}